# Crystal Facet Effect in Plasmonic Catalysis

*Yicui Kang [§ $], Simão M João[⊥ $], Rui Lin[§]\*, Li Zhu[§], Junwei Fu [Π], Weng-Chon (Max) Cheong [&],*
*Seunghoon Lee[#], Kilian Frank[¢], Bert Nickel[¢], Min Liu [Π], Johannes Lischner[⊥]\*, Emiliano*
*Cortés[§]\**

[§] Nanoinstitute Munich, Faculty of Physics, Ludwig-Maximilians-Universität München, 80539 München, Germany.

[⊥] Departments of Materials and Physics and the Thomas Young Centre for Theory and Simulation of Materials, Imperial College London, London, U.K.

[Π] Hunan Joint International Research Center for Carbon Dioxide Resource Utilization, School of Physics and Electronics, Central South University, Changsha 410083, P.R. China.

[&]Faculty of Innovation Engineering (FIE), Macau University of Science and Technology, 999078, Macau, P.R. China.

[#] Department of Chemical Engineering (BK21 FOUR Graduate Program), Dong-A University, Busan, 49315 South Korea.

[¢] Faculty of Physics and Center for Nanoscience (CeNS), Ludwig-Maximilians-Universität, Geschwister-Scholl-Platz 1, 80539 München, Germany.

[$]Contributed equally.

\*E-mails:

Rui.Lin@physik.uni-muenchen.de, j.lischner@imperial.ac.uk, Emiliano.Cortes@lmu.de

KEYWORDS: plasmonic catalysis, facet effect, carbon dioxide reduction, hot carriers, low-coordinated sites.

ABSTRACT:

In the realm of plasmonic catalytic systems, much attention has been devoted to the plasmon-derived mechanisms, yet the influence of nanoparticles' crystal facets in this type of processes has been sparsely investigated. In this work, we study the plasmon-assisted electrocatalytic $CO_2$ reduction reaction using three different shapes of plasmonic Au nanoparticles - nanocube (NC), rhombic dodecahedron (RD) and octahedron (OC) - with three different exposed facets: {100}, {110} and {111}, respectively. These particles were synthesized with similar sizes and LSPR wavelengths to reveal the role of the facet more than other contributions to the plasmon-assisted reaction. Upon plasmon excitation, Au OCs exhibited nearly a doubling in the Faradaic efficiency of CO (FE(CO)) and a remarkable threefold enhancement in the partial current density of CO (j(CO)) compared to the non-illuminated response, NCs also demonstrated an improved performance under illumination. In contrast, Au RDs showed nearly the same performance in dark or light conditions. Temperature-dependent experiments ruled out heat as the main factor in the enhanced response of Au OCs and NCs. Large-scale atomistic simulations of the nanoparticles' electronic structure and electromagnetic modeling revealed higher hot carrier abundance and electric field enhancement on Au OCs and NCs compared to RDs. Abundant hot carriers on edges facilitate molecular activation, leading to enhanced selectivity and activity. Thus, OCs with the highest edge/facet ratio exhibited the strongest enhancement in FE(CO) and j(CO) upon illumination. This observation is further supported by plasmon-assisted $H_2$ evolution reaction experiments. Our findings highlight the dominance of low coordinated sites over facets in plasmonic catalytic processes, providing valuable insights for designing more efficient catalysts for solar fuels production.

**Introduction**

Fossil fuels have been the dominant source of cheap energy for the past decades, but it excessive use has caused the release of large amounts of $CO_2$ to the atmosphere, producing harmful consequences to the planet. While a complete solution to this problem will involve the generation of renewable energy at large scale, decarbonization is a necessary and urgent step that must be taken right now. Artificial photosynthesis offers an attractive implementation to these solutions, by mimicking the natural process of reducing $CO_2$ into chemical fuels using sunlight[1]. However, the $CO_2$ reduction reaction ($CO_2$RR) is a multi-electron and multi-proton process, leading to many reaction pathways. Therefore, achieving high selectivity has always been a challenge for $CO_2$RR[2-4]. As a result, effort to improve the performance of catalysts has been widely made, including modifying the component[4-9], size[5,10], geometric structure[11-14], as well as the facets[15-17] of the materials. Among them, the crystal facet effect is regarded as a crucial factor that can affect catalytic activity and selectivity by tuning atomic arrangement[18], terrace and step edge defects[19], adsorption energy of intermediates[20], etc. Besides, distinctive facets also exhibit different physical properties, e.g. work function[21], electronic states[22], and electron mean free path[23]. Recent years have seen the wide investigation of the facet effect on heterogeneous catalysts in the field of thermocatalysis[24-26], electrocatalysis[18,19,27-37], photocatalysis[38,39] and photo-electrocatalysis[40,41]. For example, Hori et al. focused on the facet effect of single crystal Pt in electrocatalytic $CO_2$RR system back in 1995. They confirmed that Pt (110) has a more than 10 times rate of CO formation than Pt (111)[37], demonstrating the big potential of modifying the exposed facets in regulating the selectivity and activity of catalysts for $CO_2$RR.

In parallel, the emergence of plasmonic catalysis as a novel field has garnered significant attention due to its distinct properties and potential for enhancing activity and selectivity in diverse catalytic processes. Plasmonic catalysis involves the excitation of localized surface plasmons (LSP), which refers to the resonant oscillation of conduction electrons induced by photons[6]. At the LSP resonance (LSPR), nanoparticles (NPs) concentrate electromagnetic fields on their surfaces, leading to a significant field enhancement. Subsequently, the surface plasmons decay into energetic electron and hole pairs through Landau damping. These carriers with high energy can then transfer into unoccupied levels of acceptor molecules adsorbed on the catalyst surface and induce chemical transformation[42,43]. Besides, the chemical bonds of adsorbates are selectively activated by plasmonic catalysts in direct electron transfer processes, introducing the possibility of plasmon prompting specific chemical pathways selectively[43-45]. Additionally, relaxation processes with the lattice can induce local heating, further enhancing the catalytic performance[46,47]. As such, plasmonic catalysis of many reaction regimes has been studied, such as $H_2$ dissociation[48,49], $O_2$ activation[50], $N\equiv N$ dissociation[51], $H_2$ evolution[52,53], as well as $CO_2RR$[54]. These studies demonstrate the capability of plasmonic catalysis to enhance chemical reactions, modify selectivity, and even induce novel reaction pathways[7,55,56]. Moreover, the LSP offers high tunability and presents opportunities to manipulate light absorption at nanometer and femtosecond scales[57]. This remarkable feature positions plasmonic catalysis as a promising avenue for enhancing the efficiency and selectivity of solar energy conversion, thereby attracting significant attention[58]. Yet the experimental evidence of plasmons improving the selectivity in electrocatalytic $CO_2RR$ system and the study of the underlying mechanism are limited. Hence, the primary objective of this project is to study the facet effect in plasmonic $CO_2RR$, building on previous evidence of its significance in thermal catalysis, electrocatalysis, photocatalysis, and related areas.

For this purpose, Au is a suitable material choice due to its favorable catalytic activity for $CO_2$RR and strong surface plasmon resonance response in visible regions.

Based on the fundamentals mentioned above, we synthesized 3 morphologies of Au nanoparticles (NPs): nanocubes (NCs), rhombic dodecahedrons (RDs) and octahedrons (OCs) - with the same phase, similar size, but distinctive exposed facets. The electrocatalytic $CO_2$RR selectivity on three Au NPs differ significantly in between them, which originate from the distinct $CO_2$ activation energy of their various exposed facets, as confirmed by the DFT calculations. However, when introducing plasmons into the system, Au OCs and NCs exhibit significant increases in the selectivity to CO, while RDs only show bare improvement. Then large-scale atomistic simulations and electric field modeling were conducted to investigate the underlying mechanism. Our findings reveal that the response to plasmons in catalytic processes predominantly relies on the amount and spatial distribution of plasmon-induced hot carriers. In this context, edges are pivotal, while the role of facets appears to be insignificant. These results shed a light for designing catalysts for more efficient $CO_2$ reduction and sunlight utilization for decarbonization.

## Results

### Synthesis and characterization of Au NPs

Gold nanocrystals with three different exposed facets Au {100}, {110} and {111} --- were prepared by seed-mediated growth approaches in the presence of CTAB surfactant[59,60], they were in the morphologies of nanocubes (NCs), rhombic dodecahedra (RDs) and octahedra (OCs), respectively. The detailed synthetic process is described in Methods section. SEM and TEM images of the synthesized single-crystalline Au nanocrystals with monodispersed shapes and sizes are shown in Figure 1(a-c) and Figure S1, respectively. The Au NCs, RDs and OCs have average sizes of around 60 nm (Figure S2). The face-centered cubic (FCC) phase of all three Au nanoparticles was confirmed by X-ray diffraction (XRD), as evidenced by the presence of characteristic peaks of (002), (020), (200), (202), and (220) (Figure 1g). To gain further insights into the structure of Au nanoparticles, high-resolution transmission electron microscopy (HRTEM) and selected area electron diffraction (SAED) were performed, as depicted in Figure 1(d-f). The observed d-spacing values in HRTEM images of Au NCs, RDs, and OCs were 0.209, 0.141, and 0.242 nm, respectively. These values align with the surface termination of Au {100}, {110}, and {111} facets in the face-centered cubic (FCC) phase, as presented in Table S1. The distinctly sharp and well-defined spots in SAED patterns confirmed the single-crystalline structure of Au NPs. These findings are also consistent with existing literature on nanoparticle characterization[59]. UV–vis–NIR spectroscopy was employed to examine the optical properties of the Au NPs. Figure 1(h) illustrates that all three samples exhibit extinction spectra with a single peak at around 540 nm. This indicates the presence of a single LSPR mode in each sample, with similar excitation wavelength. The sharpness of the peaks further confirms the uniform sizes and low size-dispersions for all synthetized Au nanoparticles, consistent with the observations from SEM and HRTEM.

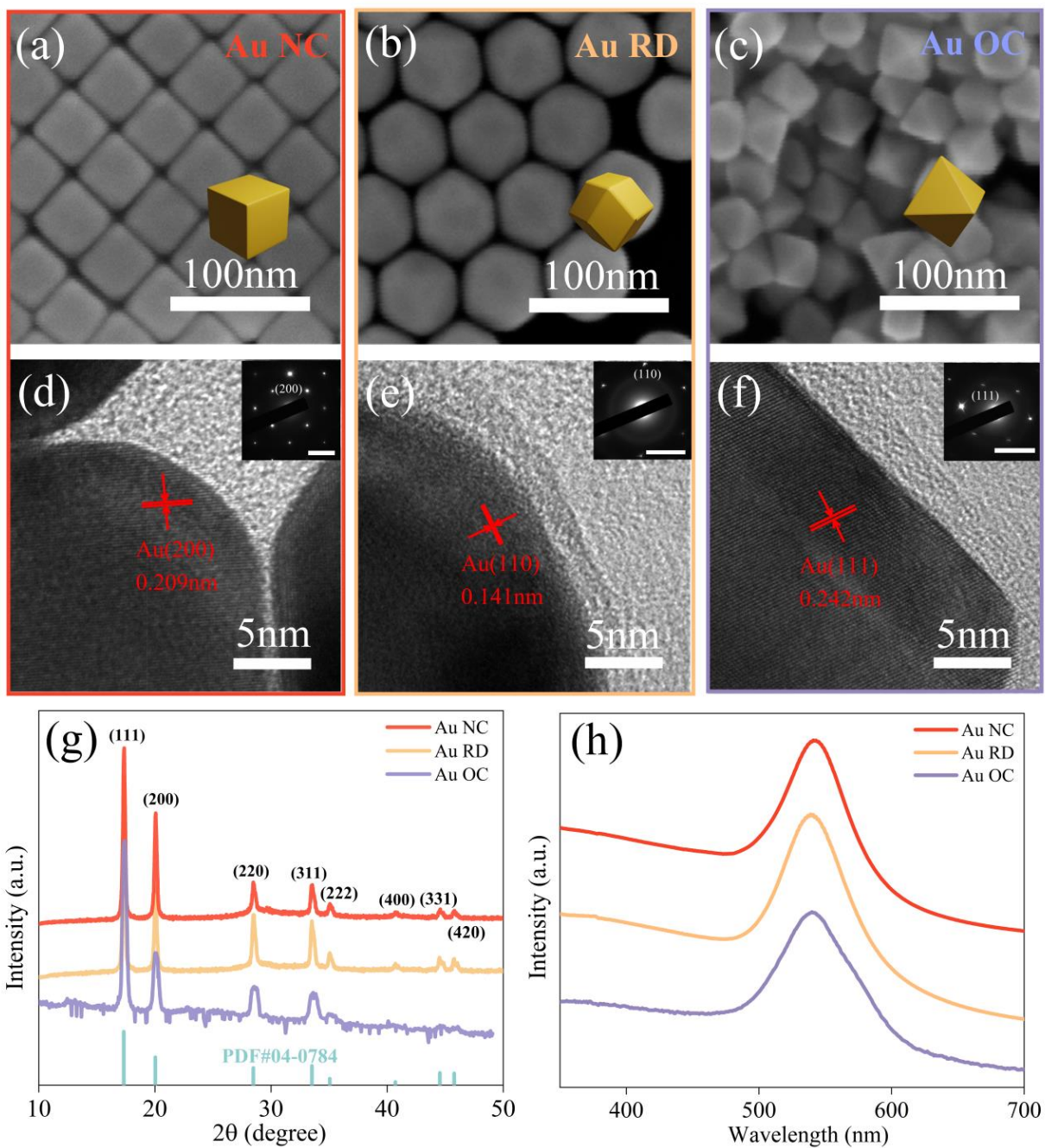

**Figure. 1** (a-c) SEM images of (a) Au NCs, (b) RDs and (c) OCs. (d-f) HRTEM and SAED (inset) patterns of (d) Au NCs (e) RDs and (f) OCs, the scale bars in insets are 5 1/nm. The average distances between fringes and the corresponding Au facets are marked in red. (g) XRD spectrums of Au NCs, RDs, OCs and XRD ruler of Au with Mo as target for fcc phase. (h) UV-vis spectrum of Au NCs, RDs and OCs with absorption peaks at 543 nm, 539 nm and 542 nm, respectively.

**Electrocatalytic performance of Au NPs**

As shown above, we have synthesized three uniform Au nanostructures NC {100}, RDs

{110} and OCs {111} with the same phase, similar sizes and LSPR wavelengths. $CO_2RR$ is

considered a promising solution for reducing excessive $CO_2$ in the atmosphere. However, the

major hurdles in $CO_2RR$ lie in its high activation energy and limited selectivity. Recognizing the

potential of Au as a catalyst with low activation energy and high selectivity, we conducted

measurements to evaluate the electrocatalytic performance of Au NPs with three different exposed

facets (for the details of our experiments see SI, Scheme S1, Figure S3 and S4). The corresponding

current-time (I-t) curves and linear sweep voltammetry (LSV) curves are presented in Figure S5

and Figure S6, respectively. Under all applied potentials, the major gas products detected were carbon monoxide and hydrogen. Liquid products were traced by proton nuclear magnetic resonance ([1]H NMR) and no products were detected (see Figure S7). Control experiments with carbon powder deposited on carbon paper were also conducted to confirm that $CO_2RR$ is catalyzed by Au NPs instead of carbon, as shown in Figure S8.

The FE(CO), FE(H$_2$) and the corresponding partial current density of Au NCs, RDs and OCs

are shown in Figure 2a and S9. The electrocatalytic $CO_2RR$ selectivity of three Au NPs differ

significantly in between them. Au RDs with {110} facet demonstrated the highest FE(CO) at all

applied potentials from -0.53 to -0.88 $V_{RHE}$. Notably, at -0.67 $V_{RHE}$, RDs achieved the highest

FE(CO) of 94%. In comparison, NCs {100} presented a lower CO selectivity (69%) and OCs

{111} showed the lowest among three Au NPs with 51% FE(CO) at -0.67 $V_{RHE}$.

DFT calculations were performed to elucidate the reasons behind the different electrocatalytic $CO_2$ reduction performance among the three Au nanoparticles. The calculation details are described in SI. The calculations revealed that the activation of $CO_2$ is the rate-determining step, as depicted in Figure S10. The formation energy of COOH* on Au{110} (RDs) is 1.1 eV, which is lower than 1.2 eV on Au{100} (NCs) and 1.4 eV on Au{111} (OCs). The lowest COOH* formation energy on Au{110} (RDs) facilitates the $CO_2$ molecule activation and the CO generation. In summary, regarding FE(CO), the electrocatalytic performance of Au NPs with different exposed facets can be ordered as Au RD {110} > NC {100} > OC {111}, which is a consequence of the different energy barriers observed on the corresponding facets. The results are also consistent with previous works based on different Au facets[18].

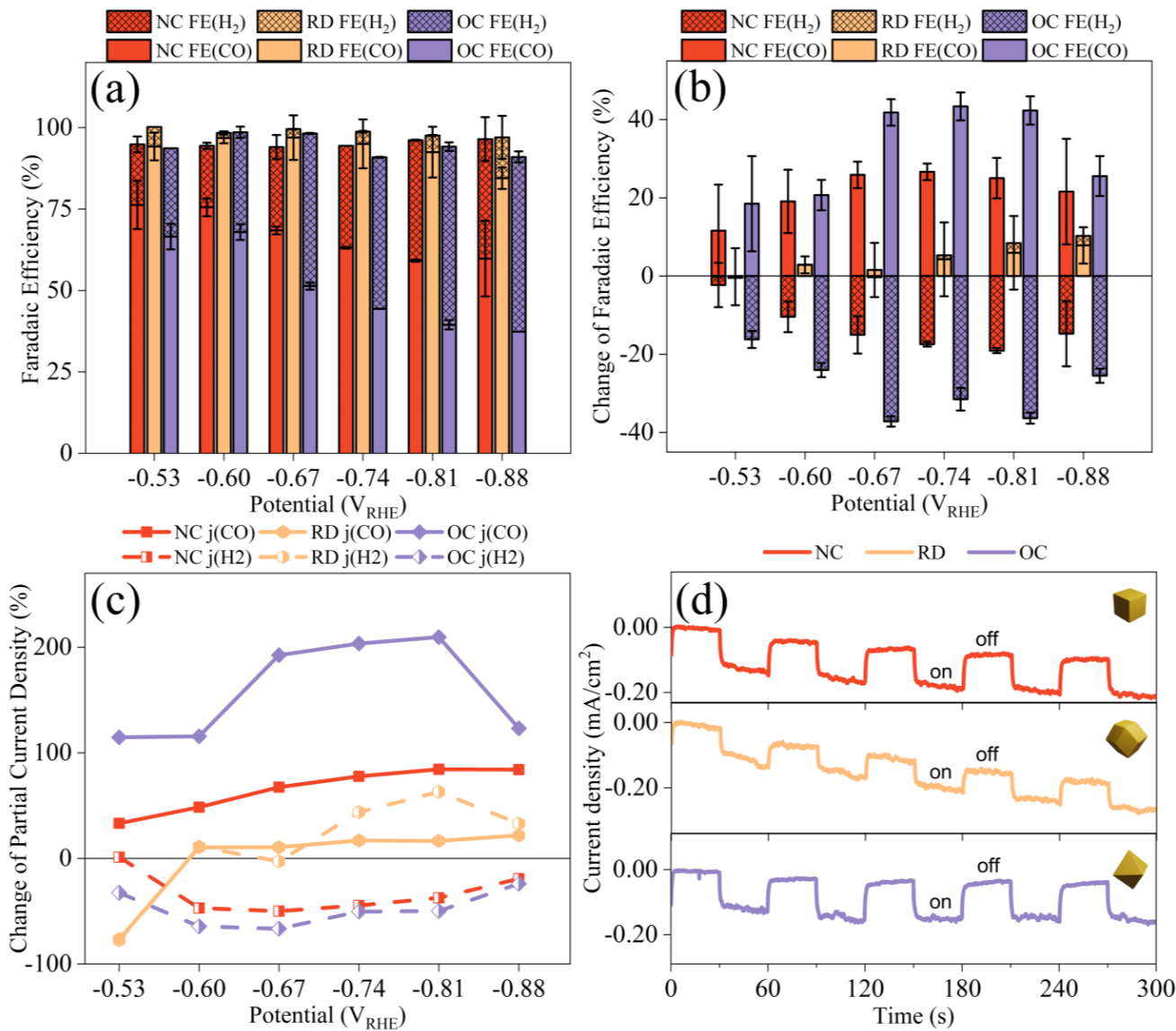

**Figure. 2** (a) Faradaic efficiencies (FE) of CO and $H_2$ production on Au NCs (red), RDs (yellow) and OCs (purple) in electrocatalytic $CO_2$ reduction system. (b) Change in the absolute value of FE for CO and $H_2$ [$FE_{light}$ - $FE_{dark}$] on Au NPs when illuminated by 525nm LED. (c) Percentage change in partial current density for CO and $H_2$ [$(j_{light}$-$j_{dark})$/$j_{dark} \times 100\%$] on Au NPs when illuminated. (d) Chopped-light chronoamperometry with 525 nm LED at -0.81 $V_{RHE}$ on Au NPs.

**Plasmon-assisted electrocatalytic performance of Au NPs with different exposed crystal facets**

As mentioned above, plasmon excitation has recently emerged as a new way of exciting photoactive materials that may help improving the electrocatalytic selectivity and current density of a wide range of reactions[56]. To investigate how plasmons affect our system, the plasmon-assisted electrocatalytic $CO_2RR$ on Au NPs was subsequently studied. A Light-emitting diode (LED) with a wavelength of 525 nm was chosen as the illumination source because this wavelength is close to the LSPR of Au NPs. The cathode was continuously illuminated by the LED ($610mW/cm^2$) in this system. The corresponding I-t and LSV curves are presented in Figure S5 and Figure S6, respectively. Similar to the electrocatalytic $CO_2RR$, the major gaseous products in plasmon-assisted system were also CO and $H_2$, and no liquid products were detected by $^1H$ NMR, as shown in Figure S7. Control experiments with carbon powder were also done to eliminate the influence of carbon, as shown in Figure S8.

Plasmons showed a significant impact on the catalytic selectivity and activity of Au NPs, as shown in Figure 2b and 2c. While Au OCs display low FE(CO) in electrocatalytic $CO_2RR$, the excitation of plasmons notably enhanced their selectivity. Illumination resulted in an increase of at least 19% in the absolute value of FE(CO) on Au OCs compared to pure electrocatalysis at all tested potentials. Notably, the most dramatic increase of FE(CO) was 43%, taking place at −0.81 $V_{RHE}$, where it increased from 44% in dark conditions to 87% with illumination, which was twice as high as that without plasmonic excitation. When illuminated, Au NCs also showed a FE(CO) enhancement between 12% to 27% at all applied potentials. In contrast, Au RDs, which had the highest FE(CO) in pure electrocatalytic $CO_2RR$, showed minimal improvement when they were illuminated. More specifically, under most potentials the increases of FE(CO) were less than 5% on Au RDs. Furthermore, the partial current density of CO (j(CO)) on Au OCs and NCs also showed noticeable increase with illumination as shown in Figure 2c. Specifically, at -0.81 $V_{RHE}$, the j(CO) on Au OCs experienced a remarkable enhancement of 210% with plasmon excitation compared to electrocatalysis alone. Similarly, Au NCs displayed an improvement of 84% at this potential upon introducing plasmons. However, Au RDs only exhibited a modest increase of 16% in the presence of light. The increase in CO partial current density on Au OCs and NCs resulted from two factors: the enhancement of FE(CO) and the large photocurrent generated by the hot carriers produced upon the LSPR decay. Figure 2d presents the chopped I–t curves of three Au NPs under illumination of 525 nm. Au OCs and NCs showed pronounced large plasmon-induced photocurrents of 0.12 and 0.11 mA/cm$^2$, respectively, indicating their high photocurrent responses. In contrast, on Au RDs, LSPR only contributed with a photocurrent of 0.08 mA/cm$^2$. In summary, the introduction of plasmons in the electrocatalytic CO$_2$RR system led to substantial enhancements in activity and selectivity to CO, both Au OCs and NCs demonstrate substantial improvements. However, Au RDs, despite possessing high intrinsic electrocatalytic performance, exhibited a poor response to light and show only minimal improvement.

Then the influence of light wavelength on plasmonic catalytic performance was also examined in this study. To investigate this, we employed an LED with wavelength of 405nm operated at the same power as the 525nm LED (610 mW/cm$^2$). Despite the higher energy of 405nm photons compared to 525nm photons, the enhancement observed in FE was smaller, particularly evident in the case of Au NCs (see Figure S11). These findings provide further evidence of the remarkable efficiency of plasmon involvement in catalytic processes and its potential for enhancing catalyst selectivity.

**Study of the mechanism behind plasmon-enhanced electrocatalytic CO$_2$RR**

The underlying mechanism of the impact of plasmon excitation on distinctive Au NPs regarding electrocatalytic systems was then studied. When the LSPR is triggered, the plasmon decays nonradiatively, resulting in the generation of energetic "hot" electrons and holes[61-63]. These hot carriers can then be transferred to the reacting molecules, facilitating the catalytic process. Additionally, the energy carried by the hot carriers produced during plasmon decay can also be transferred to other electrons through electron-electron collisions, to the lattice via electron-phonon interactions, and eventually to the surroundings, leading to an increase in the surface temperature[64].

Given the exponential increase of reaction rates when increasing the temperature (Arrhenius law), we first focused on investigating the impact of heat. To explore this, we conducted experiments on the CO$_2$RR catalytic system in the dark under three different temperature conditions: room temperature (23°C), 30°C, and 40°C. We achieved these temperatures by means of external heating with a hot plate. In contrast to the scenario when plasmons are excited, the temperature-dependent experiments revealed a distinct inverse relationship regarding FE(CO), as depicted in Figure 3(a-c). As the temperature increased, there was a noticeable decline in selectivity towards CO, favoring the HER at higher temperatures. For instance, at -0.67 V$_{RHE}$, the FE(CO) on Au RDs decreased from 94% at room temperature to 51% at 30°C, and further dropped to 13% at 40°C. This trend was consistent across all three Au NPs, indicating a clear reduction in selectivity towards CO with higher temperatures. Hence, while local heating plays a significant role in various LSPR-involved systems, it is not the primary factor contributing to the observed enhancement in selectivity towards CO in this study. This finding highlights the unique properties and advantages of plasmonic catalysis compared to conventional thermal catalytic processes, as the enhancement of selectivity to CO cannot be achieved thermally[65,66].

As heat is not responsible for the observed effects under illumination, our study then focused on the impact of hot carriers generated through plasmon decay in this system. Due to the high energy of hot carriers and their ability to transfer to absorbed molecules on the surface, they play a critical role in many plasmonic catalytic processes and have been extensively investigated. The sensibility of hot carriers on the geometry of the material is known to lead to varied catalytic activities[67,68]. Hot carriers can induce processes like assisted-desorption of CO from the surface, enabling further $CO_2$ molecules to adsorb on CO-active sites, leading to an increase in current and potentially also in the FE(CO). They could also participate in other processes, such as the activation of adsorbed $CO_2$. Therefore, to explore the potential effects of hot carriers in our system, we calculated the hot-carrier generation rate in Au NCs, OCs and RDs. To accomplish this, we employed the recently developed approach of Jin and coworkers[69]. In this method, the electric potential induced by the light is first calculated using the quasistatic approximation and then used to evaluate Fermi's golden rule (FGR). The FGR is evaluated using large-scale atomistic tight-binding simulations and each nanoparticle consists of approximately 200,000 Au atoms (see Methods for more details). Figure 3(d-f) shows the distribution of hot electrons and hot holes, exhibiting pronounced peaks associated with interband transitions from states with d-band character to states with sp-band character. In addition, smaller peaks (most clearly visible for the Au NCs) are caused by intraband transitions involving initial and final states with sp-band character. These peaks arise from surface-enabled transitions, which get magnified due to the field field enhancement at the surface of the nanoparticle. The results demonstrate that Au RDs generate fewer hot carriers compared to Au NCs and OCs, while the latter two exhibit similar levels of hot carrier generation. Notably, the hot carrier generation is strongly dependent on the electric field enhancement on the surface of plasmonic nanomaterials[67]. Moreover, the direct transfer of hot electrons generated by LSPR is strongly influenced by the local electric field conditions[70]. Consequently, we conducted an analysis of the electric field distribution near the surface of Au NPs, as depicted in Figure 3(g-i). Notably, Au OCs and NCs exhibit more prominent electric field enhancements on their edges and corners compared to Au RDs. Large-scale atomistic simulations and electromagnetic modeling, in conjunction with experimental evidence of weaker light response in RDs compared to NCs and OCs, reinforce the significance of hot carriers and electric field enhancement in plasmonic catalytic processes.

Interestingly, facets can mostly account for the dark electrocatalytic performance of Au OCs, NCs and RDs in the $CO_2$RR. However, uncoordinated sites (edges) are the main responsible sites for the plasmonic-assisted $CO_2$RR in these systems. This is a remarkable result, as it highlights that plasmonic catalytic sites (or reactive hot spots) may differ from the inherent surface expected sites in the dark[46,71].

The plasmon-induced electric field enhancement was proved to be highly dependent on the nanostructure[42]. Remarkably, on all Au NPs, pronounced enhancements were observed at low-coordinated sites, specifically corners and edges, rather than facets. This is consistent with previous investigations into the atomic structure of plasmonic nanoparticles, which revealed a non-uniform spatial distribution of hot electrons: they tend to concentrate more prominently at lower-coordinated sites such as edges, as opposed to higher-coordinated sites such as facets[72]. A recent study applying single-particle electron energy loss spectroscopy also revealed the spatially inhomogeneous carrier extraction efficiency in plasmonic nanostructures[68]. Notably, investigations into the impact of low-coordinated sites have established their superiority over facets in catalytic processes[73]. Studies on the geometric properties of plasmonic nanomaterials also indicate that on the edges, the binding energy of the intermediate *COOH is lower than on facets, and binding to CO is weaker than at corners, making them more selective for $CO_2$ reduction to CO rather than $H_2$ production[74,75]. Consequently, edges with low coordination number serve as the pivotal juncture where highly active $CO_2$RR sites coincide with a high electric field and a substantial population of hot electrons. The efficient transfer of hot electrons to the adsorbed $CO_2$ molecule facilitates its activation, resulting in a significant enhancement in the activity and selectivity of $CO_2$RR. Even though these sites are also present in the dark experiments, our results indicate that they become much more active – even dominating the overall response of the system - when the plasmon is excited. The mechanism is depicted in Figure 4a. As such, Au NPs with a higher edge-to-facet ratio, such as Au OCs (as supported by calculations and specific values provided in Table S4), are expected to exhibit the most pronounced plasmonic response regarding $CO_2$RR, as corroborated by experimental observations.

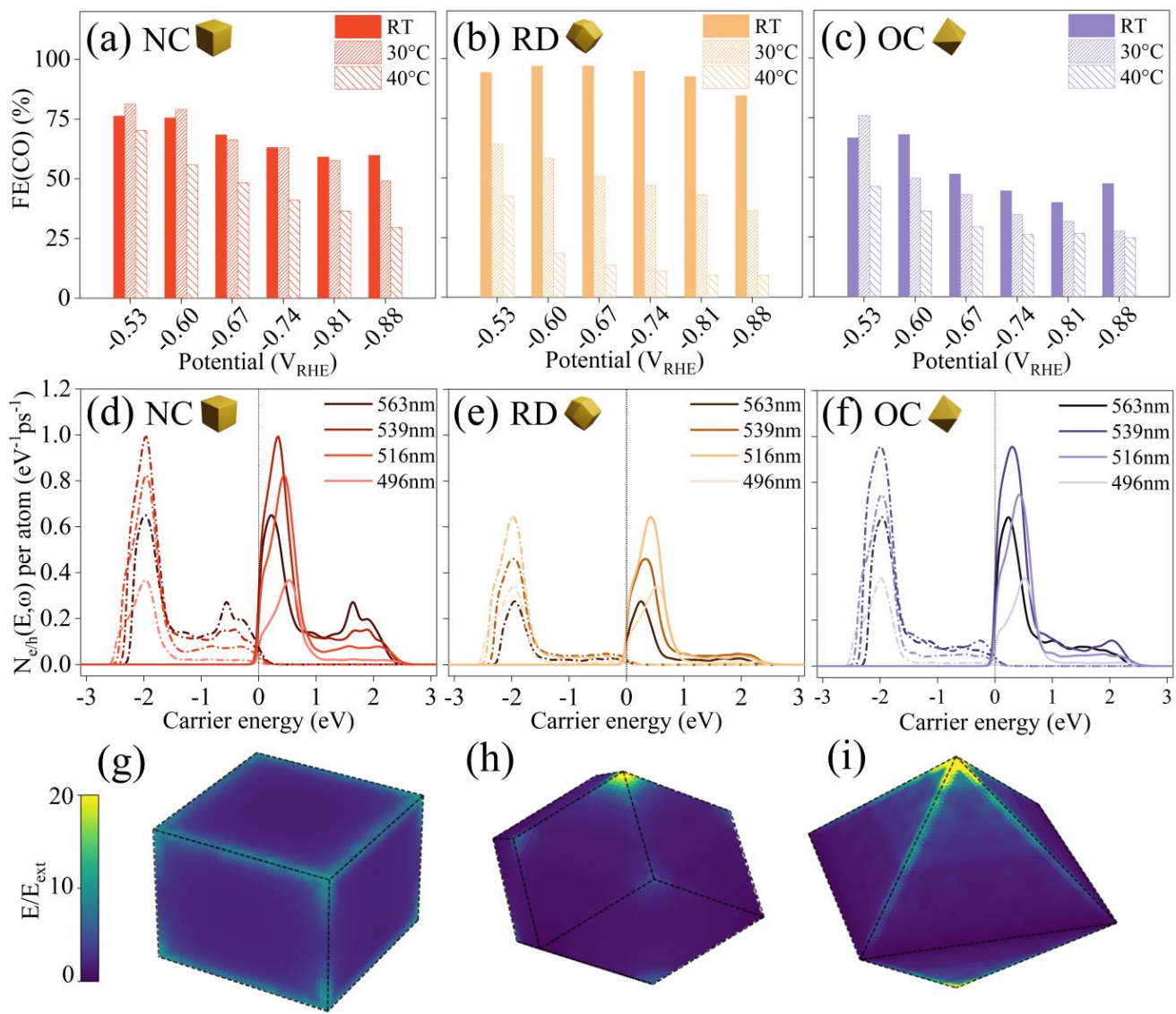

**Figure. 3** (a-c) Faradaic efficiencies (FE) for CO production at different applied potentials on (a) Au NCs, (b) RDs and (c) OCs at 20°C, 30°C and 40°C. (d-f) Hot electron (solid) and hole (dashed) generation rate for each of the (d) Au NCs, (e) RDs and (f) OCs at different electric field wavelengths. The Fermi energy is set to zero. (g-i) Absolute electric field profile in (g) Au NCs, (h) RDs and (i) OCs at the corresponding LSPR frequencies, in reference to the external applied electric field. All simulated nanoparticles have approximately 200 000 atoms.

To verify the validity of our hypothesis regarding the prominence of low-coordinated sites with high electric field and a substantial quantity of hot electrons over facets in diverse plasmonic catalytic systems, we conducted experiments utilizing the same Au NPs for the hydrogen evolution reaction (HER) process (details of experiments see SI and Scheme S2). Under dark conditions, the HER onset potentials on Au NPs are in the order of: RD{110} < NC{100} < OC{111}, as shown in Figure 4(b-d), which is consistent with the DFT simulation results (Figure S12). Upon exciting plasmons in the HER system, hot carriers generated by plasmon decay can activate the adsorbed $H_2O$ molecule and lower the onset potential. With illumination, Au OCs and NCs, which possess a large number of hot carriers and exhibit high electric field enhancement as shown in Figure 3, show a remarkable reduction in HER onset potential by 0.033 $V_{RHE}$ and 0.021 $V_{RHE}$, respectively. In contrast, Au RDs only present a slight improvement of 0.007 $V_{RHE}$ (Figure 4(b-d)). The remarkable plasmon interaction observed in Au OCs and NCs, coupled with the limited response in HER exhibited by Au RDs, provides additional support for the significance of abundant hot carriers and intense electric field enhancement in plasmonic catalytic systems. The largest reduction in onset potential observed in OCs, which possess a high edge/facet ratio, again emphasize the dominance of low-coordinated sites over facets in plasmonic catalytic processes.

In summary, contrary to electrocatalytic systems, the role of facets is not prominent in plasmonic catalytic systems. Instead, low-coordinated sites such as edges with significant electric field enhancements, along with a high distribution of hot electrons play more significant roles. These results, which deviate from conventional catalytic processes, exhibit a unique signature of plasmonic catalysis. This signifies the potential of plasmonic catalysis to achieve novel functionalities that are not attainable by other catalytic processes.

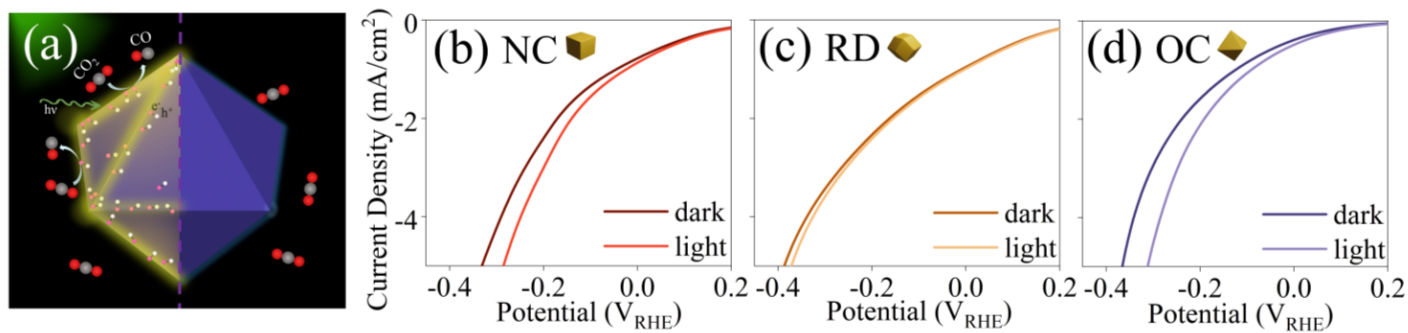

**Figure. 4** (a) Schematic diagram of plasmon enhancing the activity and selectivity of $CO_2RR$ on Au NPs. (b-d) HER performance on Au NPs. LSV curves of (b) Au NCs, (c) Au RDs and (d) Au OCs in dark and with 525nm illumination.

## Discussion

We studied the effect of plasmon excitation in electrocatalytic $CO_2RR$ and HER systems based on three Au NPs morphologies: Au NCs, RDs and OCs. In pure electrocatalytic $CO_2RR$ system, Au RDs showed the best FE(CO), with values as high as 94% at -0.67 $V_{RHE}$, NCs and OCs presented lower selectivity, with 69% and 51%, respectively. These differences primarily stem from the distinct formation energy of COOH* on the specific exposed Au facets, confirming the importance of facets in pure electrocatalysis[18]. However, when introducing plasmons into the system, Au OCs – which had the lowest CO selectivity in electrocatalytic $CO_2RR$ – presented the most significant enhancement: at -0.81 $V_{RHE}$ the FE(CO) with illumination was twice as high as the FE(CO) in dark conditions, resulting in an increase from 44% to 87%. Au NCs also showed an FE(CO) increase of 26% under the same conditions. In contrast, Au RDs demonstrated the modest increase of only 5% in FE(CO). Moreover, at this potential Au OCs exhibited a remarkable 210% enhancement of j(CO) and NCs also showed an 84% improvement, yet RDs only demonstrated a slight improvement of 16% upon plasmon excitation. These findings underscore the substantial influence of plasmons on the selectivity and activity, with distinct variations observed among different catalysts.

To further investigate the underlying mechanism behind the divergent responses to plasmonic effects, we initially examined the impact of temperature. Our control experiments clearly show that FE(CO) decreases when the temperature increases for our Au catalysts, which implies that plasmonic heating does not play a role in the improved FE(CO) performance of the Au catalysts under illumination. This highlights the existence of a different mechanism for plasmonic catalysis compared to thermal catalysis. Trying to rationalize these results, our focus shifted towards exploring the generation of hot carriers. Based on large-scale atomistic simulations of hot carrier generation and electromagnetic field modeling of the Au NPs, we found that Au OCs and NCs generated more hot carriers and stronger electric field enhancement than Au RDs. These findings align with our experimental results of stronger light response of Au OCs and NCs than RDs, underscoring the significance of hot carriers and electric field in this plasmonic electrocatalytic processes. Furthermore, electromagnetic field modeling revealed significantly stronger enhancements at low coordinated sites such as corners and edges, suggesting a higher concentration of hot carriers at these locations compared to facets. Moreover, previous studies have also demonstrated the catalytic favorability of edges in $CO_2RR$[74,75].

Based on these observations, we proposed a mechanism for the plasmon-enhanced $CO_2RR$ performance: low coordinated sites on nanostructures concentrate the electric field more intensely, resulting in a higher abundance of hot carriers. These abundant hot carriers on the edges facilitate the activation of $CO_2$ molecules, leading to enhanced selectivity and activity. In contrast, facets do not play a significant role in this process. To validate our hypothesis regarding the greater importance of low-coordinated sites compared to facets in diverse plasmonic catalytic processes, a study based on a different reaction - HER - was then carried out. Au OCs and NCs, which generated a larger number of hot carriers and were located at edge sites with significantly enhanced electric fields, exhibited notable enhancements in catalytic activity upon plasmon excitation. On the other hand, RDs with fewer hot carriers displayed a weaker plasmonic response, providing further confirmation of our proposed mechanism.

Our results show overall the effect of plasmons in electrocatalytic $CO_2RR$ systems based on three morphologies of Au NPs and reveals the significance of low-coordinated sites over facets in plasmonic catalytic processes. This study provides – from an atomistic perspective – valuable insights into the non-thermal properties of plasmonic catalysis and offers valuable guidance for the design of the next generation of plasmon-based catalysts.

**Methods**

**Synthesis of Au NPs.**

1. Au NCs. Au NCs and Au RDs were synthesized in accordance with literature[59]. The materials for synthesis see SI. Firstly, 0.6 mL 10 mM ice-cold $NaBH_4$ was rapidly injected to 10 mL pre-prepare aqueous solution containing 0.25 mM $HAuCl_4$ and 75 mM CTAB. Then the brown solution was kept stirring slowly at 30 °C for 2 hours and diluted 100 times with DI water and used as seed hydrosol. 0.3 mL seed hydrosol was then added into 25 mL growth solution containing 0.04 mM $HAuCl_4$, 16 mM CTAB and 6 mM ascorbic acid and mixed by a vortex-mixer for 10 s. The reaction mixture was left undisturbed at 25 °C overnight and seed preparation was finished. For the growth of Au NCs, 0.150 mL of 25 mM $HAuCl_4$ was quickly added into 8 mL seed solution, and left undisturbed at 30 °C for 2 hours after being mixed by a vortex-mixer for 10 seconds. The red-purple color indicated the formation of Au NCs.

2. Au RDs. 0.5 mL of 0.1 M ascorbic acid, 0.55 mL of 0.2 mM NaOH and 0.2 mL 25 mM HAuC1$_4$ were added into 8 mL above synthesized Au NCs solution in order. The whole solution was left undisturbed at 30 °C for 2 hours till a red-brown Au RDs colloidal was obtained.

3. Au OCs. Au OCs were synthesized in accordance with literature [60]. Au octahedral seeds were firstly prepared: 0.6 mL 10 mM ice-cold NaBH$_4$ solution was injected into the mixture of 87.5 µL of 20 mM HAuCl$_4$ solution and 7 mL of 75 mM CTAB solution. After 3 hours of gentle stirring, the mixture was diluted 100-fold with DI water to get the octahedral seed hydrosol. 0.15 mL seed hydrosol was added to the growth solution containing 25 µL of 20mM HAuCl$_4$, 0.387 mL of 38.8 mM ascorbic acid and 12.1 mL of 16 mM CTAB solution, and mixed by the vortex mixer thoroughly, then left unperturbed at 25 °C overnight. Next, 5 mL previous obtained solution was added into 12.5 mL of second growth solution containing 16 mM CTAB, 0.04 mM HAuCl$_4$ and 1.2 mM ascorbic acid and kept undisturbed overnight. A purple color of the solution indicated the formation of Au OCs.

**Characterization.**

TEM and HRTEM measurements were carried out on JEM1011 operated at 80kV and JEM-2100F operated at 200 kV, respectively. SEM were carried out on Ultra plus scanning electron microscope (Zeiss) with 10 kV beam intensity. UV–vis–NIR absorption spectra were measured on UV/VIS/NR spectrometer Lambda 750 (Perkin Elmer). XRD was carried out by Kilian Frank with custom built Molybdenum Kalpha X-ray reflectometer/diffractometer from physics department of LMU. The samples were measured at a fixed incidence angle of 10°, and intensity at different scattering angles (2 theta) were recorded from 10 to 60° in 2500 steps (0.02°/step) for 10s each using a point detector (NaI scintillation counter). EA (C, H, N) test was measured with Heraeus Elementar Vario EL instrument.

**TDDFT calculation.**

**1. Details on the numerical calculation of the hot carrier generation rate.**

The large-scale atomistic hot carrier generation simulations were carried out following a recent two-step method developed by Jin et al[69] :The first step is to calculate the electric field inside the NP and the second step is to use this information to get the hot carrier generation rate using Fermi's Golden Rule (FGR).

For the first step, the NP is considered as a dielectric under an applied external electric field. The nanoparticle is assumed to behave as a dielectric in vacuum with dielectric constant $\epsilon$ ($\omega$) for an electric field of frequency $\omega$. The values of $\epsilon$ ($\omega$) are obtained from experimental data of bulk gold[76]. Within the quasistatic approximation, the electric field inside the nanoparticle can be found by solving Laplace's equation for the electric potential using commercially available software such as COMSOL and imposing the boundary conditions that the electric field at infinity is uniform.

Figure 3 (g-i) shows the absolute value of the electric field in each of the nanoparticle geometries, with the incident electric field pointing in the z direction and frequency tuned to the LSPR. The nanoparticle sizes were chosen such that each contains around 200, 000 gold atoms, lying comfortably inside the range of sizes required for the quasistatic approximation to hold. Field enhancement effects are clearly visible at the corners and edges of all nanoparticle geometries.

Once the electric potential has been found, it acts as the perturbation driving the system out of equilibrium. For the second step, the rate at which electron-hole pairs are generated is found using the FGR which is evaluated using the Kernel Polynomial Method (KPM)[77] to avoid explicit calculation of eigenfunctions and eigenenergies (see SI for more details). The quantum-mechanical properties of the electrons are described by a tight-binding Hamiltonian obtained via a two-center Slater-Koster parametrization of bulk gold[78]. This is a 9-orbital model, taking into account the 5d, 6s and 6p orbitals, while spin is only taken into account as a degeneracy factor. The atomistic NP is constructed by specifying the bounding facet planes and filling the inside with a face-centered cubic lattice compatible with those atomic planes. This was the approach used to obtain the hot carrier generation rate.

ASSOCIATED CONTENT

**Supporting Information**.

The supporting information is available free of charge on the website. Experimental details, additional analysis of the electrochemical data and details of calculations.

AUTHOR INFORMATION

Yicui Kang: Kang.Yicui@campus.lmu.de

Simão M João: s.joao@imperial.ac.uk

Rui Lin: Rui.Lin@physik.uni-muenchen.de

Li Zhu: Zhu.Li@physik.uni-muenchen.de

Junwei Fu: fujunwei@csu.edu.cn

Weng-Chon (Max) Cheong: wccheong@must.edu.mo

Seunghoon Lee: seunghoon@dau.ac.kr

Kilian Frank: kilian.frank@physik.uni-muenchen.de

Bert Nickel: nickel@lmu.de

Min Liu: minliu@csu.edu.cn

Johannes Lischne: j.lischner@imperial.ac.uk

Emiliano Cortés: Emiliano.Cortes@lmu.de

ACKNOWLEDGMENT

The authors also acknowledge funding and support from the Deutsche Forschungsgemeinschaft (DFG, German Research Foundation) under Germany´s Excellence Strategy — EXC 2089/1–390776260 e-conversion cluster, the Bavarian program Solar Energies Go Hybrid (SolTech), the Center for NanoScience (CeNS), the European Commission through the ERC Starting Grant CATALIGHT (802989), the DAAD German Academic Exchange Center (57573042), the CSC-LMU program and the support from the Alexander von Humboldt foundation. We thank Susanne Ebert, Dr. Lars Allmendinger for the EA and [1]H NMR measurement, and Dr. Xiaguang Zhang, Dr. Giulia Tagliabue (EPFL) and Dr. Wenzheng Lu for helpful discussions. Y.K. acknowledges e-conversion (DFG) for the students exchange award program that supported her 3 months research stay at Central South University, China. S.M.J. and J.L. acknowledge funding from the Royal Society through a Royal Society University Research

Fellowship URF\R\191004. J.L. acknowledges funding from the EPSRC programme grant EP/W017075/1.